# Design and experimental study of superconducting left-handed transmission lines with tunable dispersion and improved impedance match

**E A Ovchinnikova[1], V P Koshelets[1,3], L V Filippenko[1,3], A S Averkin[1], S V Shitov[1,3], and A V Ustinov[1,2]**

[1]National University of Science and Technology (MISIS), Leninsky prospekt 4, Moscow 119049, Russia
[2]Physikalisches Institut, Karlsruhe Institute of Technology (KIT), 76131 Karlsruhe, Germany
[3]Kotel'nikov Institute of Radio Engineering and Electronics, Moscow 125009, Russia
E-mail: elhed@yandex.ru

**Abstract.** We continue detailed study of microwave properties of a superconducting left-handed tunable CPW transmission line (LHTL). The line consists of a central conductor, loaded with series of Josephson junctions as fixed inductors; the line is shunted with SQUIDs as tunable inductors. The inductance of the SQUIDs is varied in the range of 0.08-0.5 nH by applying an external dc magnetic field. The circuit is designed to have left- and right-handed transmission bands separated by a variable rejection band. At zero magnetic field, we observed only one pass-band between 8 and 10 GHz within the frequency range of 8-12 GHz. The rejection band is anticipated to appear between 10 GHz and 11 GHz by design, and it has been detected in our previous work. To solve the problem of standing waves and RF leak in measurements of our experimental 20-cell LHTL, we have designed a high-ratio (5-50 Ohm) wideband (8-11 GHz) impedance transformer integrated at the chip, along with improved sample holder. The experimental data are compared with numerical simulations.

## 1. Introduction

Metamaterials are artificial structures engineered of sub-wavelength artificial atoms, which form a homogeneous effective medium in respect to the electromagnetic waves of certain frequency range [1]. One of the most interesting and important implementation of the metamaterials is a left-handed medium. For the first time this type of materials was considered in 1967 by Veselago [2]. The medium was called "left" or "left-handed" (LH), because the electric field E, magnetic field H and the wave vector k form a left-handed system, while in ordinary media they form a right-handed system. Another important property of such materials is that the phase velocity of the LH media is opposite to both the direction of energy flow and the group velocity [3]. First experimental result was obtained in 2000 [4]. To study practicable issues of the concept, it seems feasible to implement a left-handed medium as a planar transmission line. The permittivity ε and permeability μ of such a transmission line can be modeled using distributed $L$-$C$ networks [5]. There is, however, a number of problems, for example, most of test setups assume a feed impedance of 50 Ohm, that is hard to reach at the chip. An unmatched measurement suffers from standing waves, while for a low-level signal, the dumping of standing waves with attenuators is not a suitable solution. In this work we are trying to solve this problem aiming for better accuracy of measurement of phase velocity at the chip.

## 2. Design and computer simulation

According to the simplified two-wire model, transmission line can be purely left-handed or right-handed, unless elements with frequency-dependent impedance are present within the network. The left-handed transmission line (LHTL) can be created by simply exchanging the positions of $L$ and $C$ in the two-wire model of the line [6]. However, it is not possible to create such line in practice, since a real structure always contain additional (parasitic) capacitances on shunt and (stray) inductances at main line just due to geometrical extension of any fractional element composing the line. For this reason, such LHTL will always be of mixed type, while LHTL or RHTL behavior can dominate within a certain frequency range [7]. In previous work [8], we proposed, designed and tested a transmission line, which can be tuned from RHTL to LHTL by application of a dc magnetic field. Tunable elements of this line are SQUIDs, which are very sensitive to magnetic fields. The tuning is possible via changing the inductance of the Josephson current of the SQUIDs that has been experimentally demonstrated in Ref. [9]. The Josephson inductance of the dc SQUID is given by the relation from Ref. [10]:

$$L_j = \frac{\Phi_0}{2\pi I_c \cos\varphi} , \qquad (1)$$

here $\Phi_0$ is the magnetic flux quantum, $I_c$ is the maximum (zero-field) critical current of the junction and $\varphi$ is the superconducting phase difference. The phase difference is defining the superconducting current through the junction; it correspondes to application of a non-zero magnetic field. The lumped-element equivalent circuit of a section of our transmission line is shown in Fig. 1. This section consists of two resonators, $C_1 L_{j1}$ and $C_2 L_{j2}$. Capacitance $C_1$ and inductance $L_{j2}$ define the left-handed properties of the transmission line below both resonance frequencies, while capacitance $C_2$ and inductance $L_{j1}$ contribute to the right-handed behavior above the resonances.

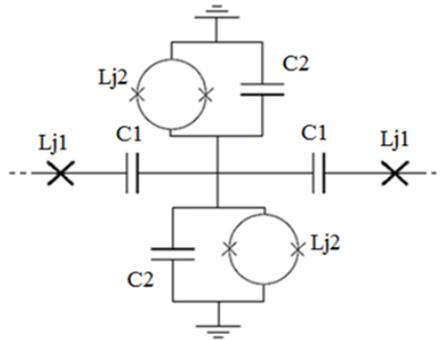

**Figure 1**. Scheme of unit cell. Crosses indicate Josephson junctions, SQUID loops and $L_j$ present the effective inductance of either SQUIDs or Josephson junctions. The experimental devices contains up to 20 cells in series that is long enough to treat the structure as a distributed one within frequency range 6-12 GHz. $C_1$= 2.1 pF, $C_2$=1.7 pF, $L_{j1}$= 80 pH, $L_{j2}$ = 80..300 pH (tunable range by magnetic field).

In previous experiments we have found at least two problems. First, the leaky sample-holder is not efficient in measuring a high rejection (low transmission) coefficients of a chip. Second, the un-matched impedance between the chip and the 50-Ohm source resulted in standing waves that heavily distort the frequency response of the circuit under measurement. It is worth to note here that accurate calibration of the RF probes with arbitrary impedance is an extremely challenging task at low temperatures.

To solve these problems and to achieve a reasonable measurement accuracy, we have designed the improved sample-holder [11] via detailed modeling with HFSS software. The holder model includes bond wires, contacts and CPW tapers integrated at the chip. These open possibility to threat the internal points of connection to the LHTL as calibrated ones.

Relaying on particular scattering matrix of the holder, we have numerically simulated resulting transmission for LHTL consisting of 20 unit cells. The simulations are shown in Fig. 2 for different values of $\varphi$ (for different magnetic fields). Due to suppressed reflections in the sample holder the transmission through the new sample holder is much smoother, and detection of low transmission of the LHTL is now more reliable. In Fig. 2a one can see that the leak is lower than 40 dB. However, the transmission through the sample can be also low, –about -30 dB in low magnetic field, Thus it is still difficult to measure the rejection band accurately..In Fig. 2b,2c the transmission is much closer to one, anticipated for bare chip that mean the holder is transparent. Once again in Fig. 2c the deep rejection cannot be seen. Checking the reason for such behavior we have found that the impedance of LHTL ranges few Ohms only. As a result, there are large reflections from standard 50-Ohm ports and standing waves interference is produced now by the LHTL itself.

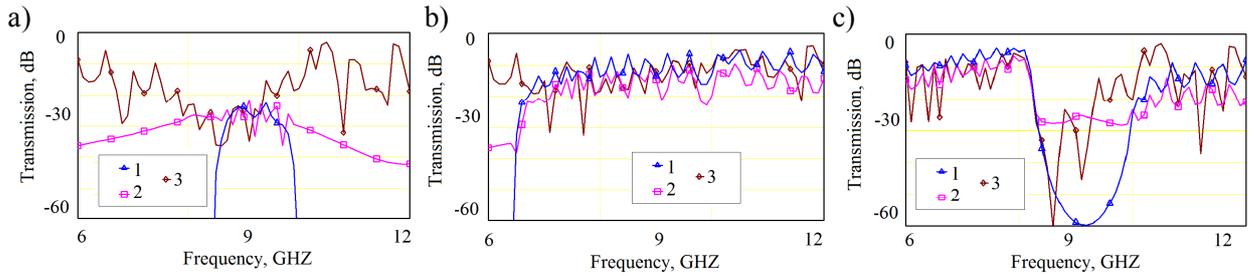

**Figure 2**. Simulated transmission $|S_{21}|$ through LHTL containing 20 unit-cells. Curves are marked: 1 – stand-alone LHTL, 2 - LHTL mounted with new sample-holder, 3 – LHTL mounted with old sample-holder. Graphs are different due to various magnetic field (different values of $\varphi$): a) $\cos \varphi = 0.8$, b) $\cos \varphi = 0.3$, c) $\cos \varphi = 0.2$.

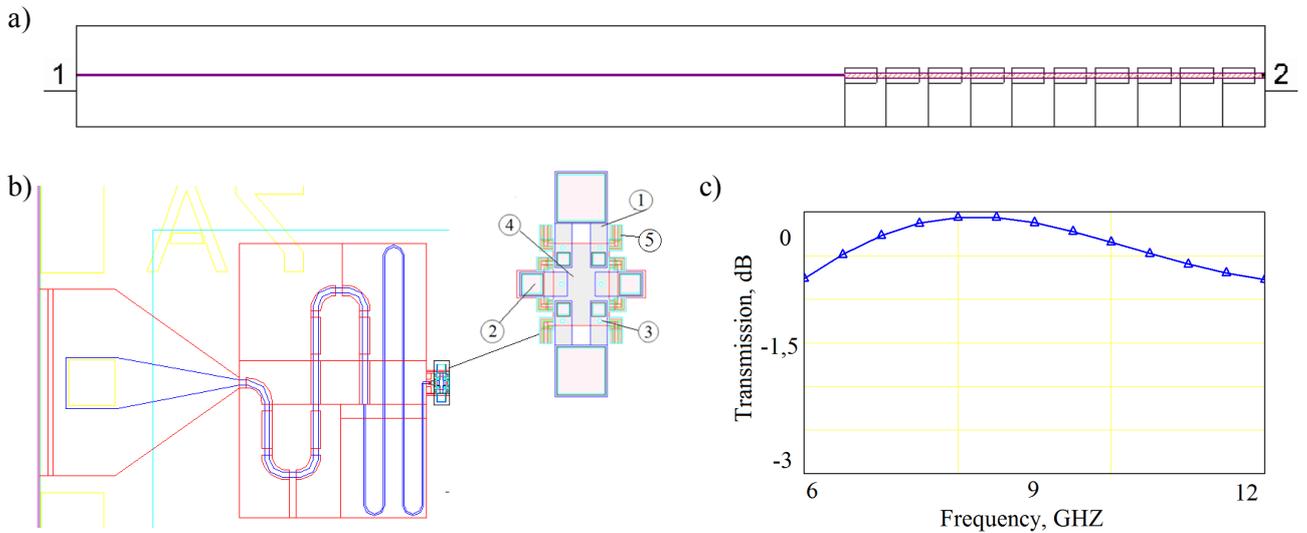

**Figure 3**. Layout of prototype 50-to-5 Ohm impedance transformer and simulation of its performance. a) Model layout of impedance transformer: the microstrip section length is 3147 µm, width is 6 µm (port 1, value 5 Ohm); the length of capacitively loaded coplanar section is 1800 µm (port 2, value 50 Ohm), width is 20 µm. b) Practicable layout of contact pads and folded transformer at the chip (left) and layout of one cell (right): 1 - SQUID loop with RF terminating capacitor to ground, 2 - via connecting neighboring cells to the left and to the right, 3 - one of six Josephson junctions of the cell, 4 - bottom electrode forming an island in the center of the cell. c) Nominal transmission of the transformer for 5 Ohm at port 1 and 50 Ohm at port 2.

To improve the impedance match, we have designed a special impedance transformer. It is known that a quarter-wave section of a transmission line with characteristic impedance $Z_0$ can compensate for reflection between two ports, $R_1$ and $R_2$, if the following condition is valid $Z_0^2 = R_1 R_2$ [12, 13]. Unfortunately, the characteristic impedance of our LHTL may change with frequency due to resonant nature of the cells and with field due to variation of inductance of SQUIDs. To relax this problem as much as possible, we have decision on trade-off and chosen a goal value $Z_0 = 5$ Ohm. Since a wide-band transformation ratio 1:10 is needed, one has to look for a multi-section transformer [14]. The series connection of two quarter-wave length transmission lines $Z_1=12$ Ohm and $Z_2=20$ Ohm can provide, according to our calculation, transmission $S_{21}$ above -1.5 dB within 6-12 GHz frequency range. To design such impedance sections, one may choose between microstrip and coplanar lines (MSL and CPW correspondingly) estimating at the same time practicable limits of available technology. There are a few limitations, which influence yield of devices fabricated with optical lithography. The strip lines should not be too long if they are narrower than 4 µm, the lines narrower than 2 µm usually are not suitable for length above 100 µm. Following the above rules, the MSL on $SiO_2$ is limited to characteristic impedance below 30 Ohm (strip must be wider than 2 µm); the CPW on Si substrate is limited to 30-100 Ohm. The first section is designed as a regular low-impedance MSL (w=6 µm), while the second high-impedance section is designed as a capacitively-loaded CPW. The integrated thin-film capacitors are added evenly along the CPW to modify permittivity of the line. As a matter of fact, such line can be treated as a metamaterial; its configuration allows for relatively independent control of both specific capacitance and specific inductance, thus providing desirable characteristic impedance and the propagation constant independent on material of CPW's substrate.

In Fig. 4 the effect of impedance transformer is demonstrated. It is clearly seen, that the transmission at low magnetic field is better for about 20 dB due to presence of the transformer. For the case of medium field, the improvement is not so large, because the impedance of the structure deviate from the design value of 5 Ohm. However, even for the deviated value of impedance, the result with transformer is essentially better that can be concluded from the smooth frequency response. To understand the transmission in more details, the impedance-frequency dependence is worth to analyze. One can see in Fig.4c that, at frequency around 8 GHz, for low impedance of LHTL the transmission is improving and for higher impedance the transmission is degrading.

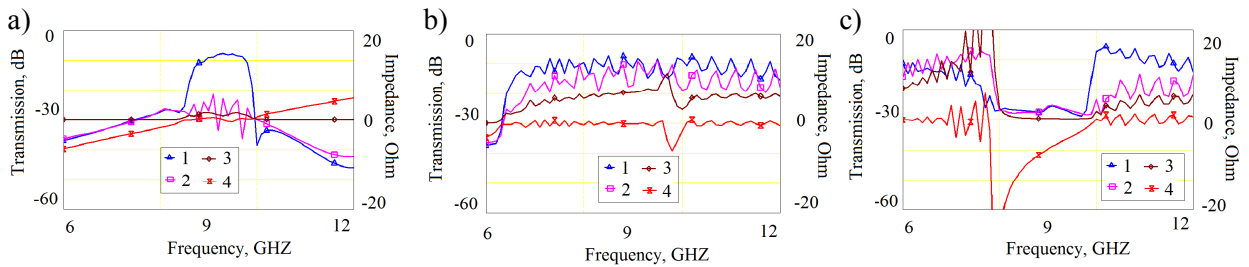

**Figure 4**. Simulated transmission $|S_{21}|$ through an array of 20 unit-cells with a new sample-holder. Curves are marked: 1 – with transformer, 2 - without transformer. 3- real impedance, 4- imaginary impedance. Graphs are different due to various magnetic field (different values of φ): a) cos φ = 0.8, b) cos φ = 0.3, c) cos φ = 0.2.

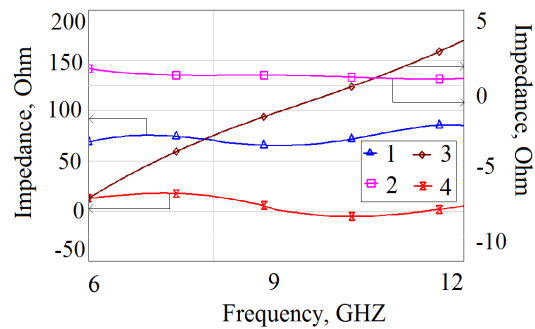

**Figure 5**. Simulation of impedance transformation: 1 – real part of impedance of holder at 50-Ohm port for 50 Ohm present at sample side (prior the transformer), 2 – real part of 50 Ohm cable at the LHTL (after transformer), 3 – imaginary part of 50-Ohm cable at the LHTL (after transformer), 4 – imaginary part of impedance of holder at 50-Ohm port for 50 Ohm present at sample side (prior the transformer).

Moreover, one can see that the best transmission occurs when the impedance is even lower than 5 Ohm. In Fig. 4a the transmission is better at impedance around 2 Ohm than in Fig. 2b for about 6 Ohm. To understand such behavior, S-parameters of the sample holder are studied. After some calculation, we may conclude that the reason can be explained with Fig. 5 showing the impedance of the holder in the range 60-80 Ohm instead of desired 50 Ohm (curve1). It is also essential, that resulting impedance at LHTL is strongly frequency-dependent. Therefore, we can now argue that the source impedance at LHTL is indeed not 5 Ohm as expected, but about 2 Ohm, and essential reactance is still present. This conclusion supports the well-known rule of "all-through design" of complex circuits, unless each chain of the system is precisely calibrated (ex. for 50 Ohm).

## 3. Experiment.

The layout of the cell, the optical image of three experimental cells and the chip wiring at the sample holder are illustrated in Fig. 6(a) and Fig. 6(b). The chip is fabricated using Nb-AlO$_x$-Nb trilayer process. The parallel plate capacitors are formed by two layers of Nb insulated with anodic oxide Nb$_2$O$_5$. Each experimental chip is 4 mm by 4 mm, and it contains two LHTL (samples) as shown in Fig. 6(b). The experimental setup is sketched in Fig. 7a. The samples are measured using the Agilent PNA-X network analyzer in a dry close-cycle cryostat at temperatures about 2 K. The signal from the network analyzer is -30 dBm, and it passes additional attenuators, 40 dB, installed at 3-K stage of the cryostat. This makes sure that the probe RF current does not exceed few percent of the critical current of the Josephson junctions. Since the output signal of the LHTL is quite weak, a low-noise coolable amplifier (LNA) is used at the 3-K stage. The magnetic field was applied by a superconducting coil external to the sample holder; the field magnitude is proportional to the current through the coil that is shown in the vertical axis of the following plots.

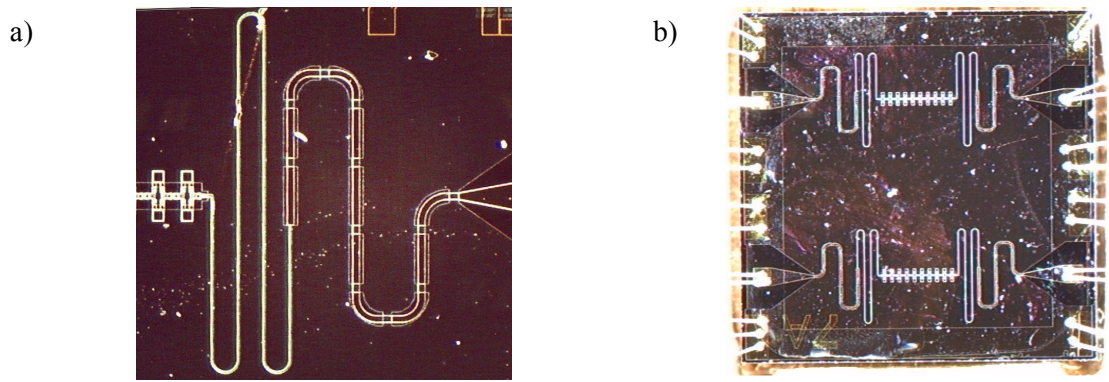

**Figure 6**. Close-up of experimental chip. a) Optical microscope image of the two-section 5-50 Ohm impedance transformer (folded structure in the center), including two cells of the LHTL (to the very left); b) Optical microscope image of the entire chip containing two identical structures (two samples); wire-bond connections of 50-Ω probe signal line to the chip are shown at the edges of the picture (to the left and to the right).

The microwave calibration is a very complex task for cryogenic measurements, as it is detailed in Ref. [15]. The calibration might be not so necessary for a well defined (flat frequency response) interface that is not the case; we need to calibrate. Extra to the holder analysis from above, a virtual calibration for the transmitted signal is attempted. To remove the background we averaged the response $S_{21}(f,H)$ and subtracted the resulting $S_{21}^{av}(f)$ from the measured data for each frequency point. It is feasible since we are interested in detecting the variations caused by the magnetic field at given frequency, not in frequency variation for a static magnetic field. The transmission through the superconducting LHTL with 10 cells is shown in Fig. 7(b); here the virtual calibration is used. The field-periodic structure in the transmission is an evidence of successful magnetic control of the LHTL.

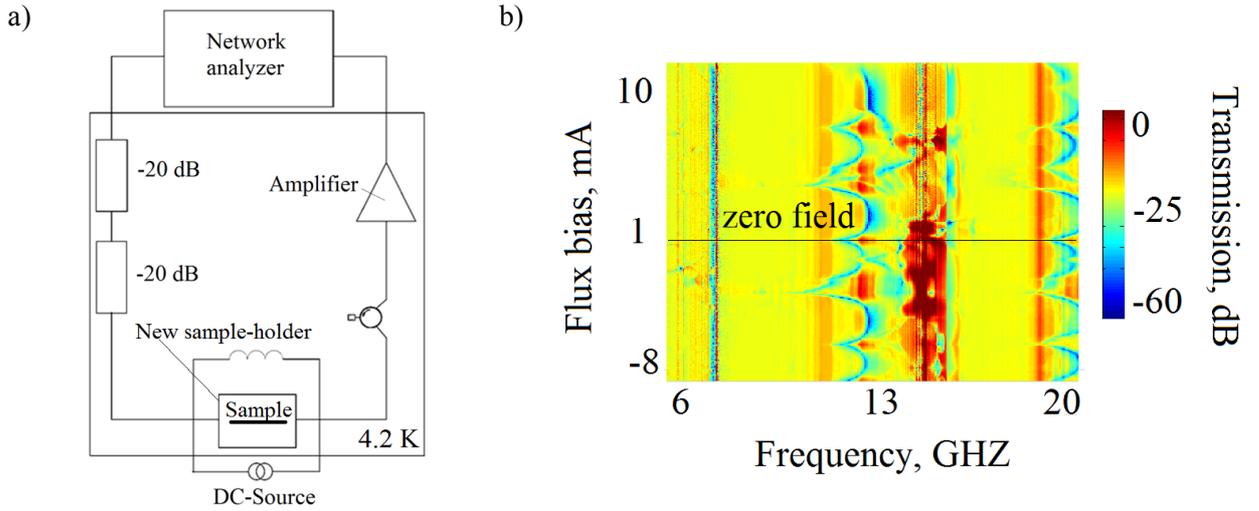

**Figure 7:** a) Simplified scheme of the experimental setup mounted within our dry 1.2-K cryocooler. The sample is placed inside a magnetic shield made of cryoperm material and equipped with internal magnetic coil. b) Quasi-3D pattern of measured transmission $S_{21}$ controled via application of magnetic field. Dark blue colors correspond to low signals, and red colors are for high signals.

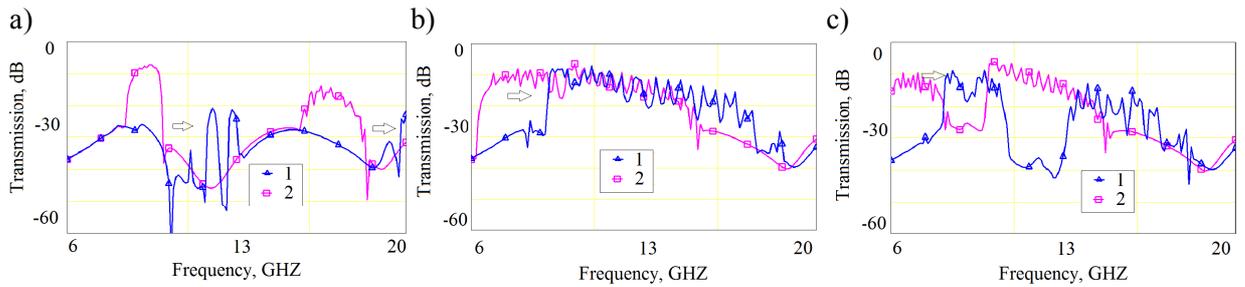

**Figure 8.** Simulated transmission $|S_{21}|$ for LHTL containing 20 unit-cells in various magnetic field (various phase φ): a) cos φ = 0.8, b) cos φ = 0.3, c) cos φ = 0.2. Curves are marked: 1 – LHTL with transformer and new sample-holder; the Josephson junctions area is larger than nominal 4 μm$^2$ (7-9 μm$^2$), 2 - LHTL with transformer and new sample-holder, area for all Josephson junctions is nominal 4 μm$^2$.

As it was predicted in computer simulation, the rejection band between the left and right transmission bands moves periodically with the field is applied as it shown in Fig. 7(b). However, the periodicity is not strict, and both of their bands are shifted at higher frequencies than expected. The most probable reason is the uneven areas of Josephson junctions that exceed the desirable fabrication tolerance. According to visual examination and test measurement, the Josephson junctions are around twice larger than nominal, and their areas are not the same. Due to larger area, larger critical current and lower Josephson inductance could be expected, resulting in higher operation frequency. This effect is presented in Fig. 7 for both bands; qualitative agreement of modeling and experiment is good. The out-of-nominal parameters of junctions lead also to worse impedance-match, because the transformer has been designed for different specific inductance of the LHTL. We have simulated LHTL with out-of-nominal parameters. In Fig. 8 one can see, that in case of low field (Fig. 8a), transmission through the line is hard to see, because it is not much higher than transmission through the sample-holder. However, one can see a thin gap in frequency around 9 GHz, that also occurs in experimental result. In higher magnetic field (Fig. 8b,c) the transmission is better and a wide gap between two frequency-range is clearer. In the experimental data one can see clearly the field-controlled rejection bands predicted in our simulation. The bands are located near 12 GHz and 19 GHz, they move and change in size with the magnetic field applied.

## 4. Conclusion.

The numerical analysis on accuracy of the LHTL measurement is confirmed experimentally. The accuracy of data is improved due to measures taken in two directions. First one is implementation of a new chip holder, which reduces input/output rf leak and allows for detection of low-transmission features of the chip. Second one is on-chip integration of 5-50 Ohm impedance transformer, which allows for essential reduction of standing waves that resulted in smoother frequency response of the entire setup. An improved sample with superconducting magnetic-field tunable left-handed transmission line (LHTL) with Josephson junctions has passed experimental verification within the wider frequency range of 6-20 GHz. Experimental data demonstrate tunability of both the transmission and rejection bands by applying a dc magnetic field. It can be seen, that parasitic transmission through the sample-holder is less, than before. However, the complexity of the circuit seems too high, and spread of parameters of Josephson junctions does not allow to see a clearly the low-field transmission band. It is worth to upgrade the LHTL design in the direction of reducing the number of junctions per cell without reducing the number of cell in the line.

**Acknowledgement.**

This work was supported in part by Ministry for Education and Science of Russian Federation with contracts 11.G34.31.0062 and K2-2014-025 (Program for Increase Competitiveness of NUST«MISiS») by the Russian Foundation of Basic Research, and by the Deutsche Forschungsgemeinschaft (DFG) and the State of Baden-Württemberg through the DFG Center for Functional Nanostructures (CFN).